\documentclass[sigconf]{acmart}
\AtBeginDocument{%
  }

\setcopyright{none}
\acmISBN{}
\acmDOI{}
\copyrightyear{}
\acmConference[Restoring Human Authenticity in AI-Mediated Communication]{CHI 2026 Workshop on Restoring Human Authenticity in AI-Mediated Communication}{April 15, 2026}{Barcelona, Spain}

\makeatletter
\renewcommand{\@mkbibcitation}{\bgroup
  \let\@vspace\@vspace@orig
  \let\@vspacer\@vspacer@orig
  \def\@pages@word{\ifnum\getrefnumber{TotPages}=1\relax page\else pages\fi}%
  \def\footnotemark{}%
  \def\\{\unskip{} \ignorespaces}%
  \def\footnote{\ClassError{\@classname}{Please do not use footnotes
      inside a \string\title{} or \string\author{} command! Use
      \string\titlenote{} or \string\authornote{} instead!}}%
  \def\@article@string{\ifx\@acmArticle\@empty{\ }\else,
    Article~\@acmArticle\ \fi}%
  \par\medskip\small\noindent{\bfseries ACM Reference Format:}\par\nobreak
  \noindent\bgroup
    \def\\{\unskip{}, \ignorespaces}\authors\egroup. \@acmYear. \@title
  \ifx\@subtitle\@empty. \else: \@subtitle. \fi
  \if@ACM@nonacm\else
    \if@ACM@journal@bibstrip
       \textit{\@journalNameShort}
       \@acmVolume, \@acmNumber \@article@string (\@acmPubDate),
       \ref{TotPages}~\@pages@word.
    \else
       In \textit{\@acmBooktitle}%
       \ifx\@acmEditors\@empty\textit{.}\else
         \andify\@acmEditors\textit{, }\@acmEditors~\@editorsAbbrev.%
       \fi
    \fi
  \fi
  \ifx\@acmDOI\@empty\else\@formatdoi{\@acmDOI}\fi
\par\egroup}
\makeatother

\usepackage{todonotes}
\usepackage{epigraph}

\begin{document}

\title[Filter Babel]{Filter Babel: The Challenge of Synthetic Media to Authenticity and Common Ground in AI-Mediated Communication}

\author{Advait Sarkar}
\affiliation{%
  \institution{Microsoft Research, University of Cambridge, University College London}
  \country{United Kingdom}
}
\email{advait@microsoft.com}

\begin{abstract}
Filter Babel is a thought experiment about a near future in which everything we read, watch, and even whom we ``meet'' is privately generated for each of us. If we each recede into a world of purely private experience, we may each develop a Wittgensteinian private language that remains intelligible to others only because an AI translator sits in the middle. This intermediation challenges the integrity of common ground and therefore of communication. On the other hand, private experience is an essential engine of identity and selfhood: as Lanier warns, one must be somebody before one can share oneself. This paper opens a discussion of the challenges and opportunities that Filter Babel might present to human communication and identity, and what constructive directions for research in AI-mediated communication might ensue.
\end{abstract}



\maketitle

\section{A Day in Maya's Filter Babel}

Maya wakes to a podcast already playing: her morning briefing, synthesised overnight from topics the system knows she likes. The host's voice is warm and familiar (though it belongs to no one). Over coffee, she scrolls through headlines, and encounters a profile of a novelist whose work uncannily mirrors her own reading history. She adores the novelist. No one else has heard of him. He was written for her alone.

On the train to work, she watches the second episode of a series she started generating last night. The protagonist, a reluctant detective, ponders a dilemma resembling a situation Maya faced last month. Across the aisle, a man laughs at his own screen. She wonders, briefly, what he's watching, but the thought dissolves before it becomes a question.

At work, her inbox contains fourteen messages, each drafted by senders' assistant agents and then re-drafted by her own assistant agent into her preferred register: direct, positive, slightly formal. The original tone, whatever it was, has been optimised away. She replies in minutes. Chains of agents will translate her replies back into dozens of forms: whatever voice she prefers to project, and whatever each recipient prefers to read (or watch, or hear). Communication is frictionless. She cannot remember the last time she was surprised by an email.

At lunch, a colleague mentions a documentary about the '96 Everest disaster. Maya nods politely. She has seen a documentary about it, as part of her 2019 season, and it featured a narration by someone who sounded like her favourite voice actor. Were they the same film? She doesn't ask. There is no reason to. Her version was made for her.

In the afternoon, she joins a discussion with three colleagues. Their agent proxies have already summarised positions, identified discrepancies, listed potential compromises, and drafted an agenda for the negotiation. The meeting is efficient. Maya speaks. The others speak. The proxies listen.

That evening, she declines a dinner invitation. There's a new episode waiting, and besides, she has nothing to talk about. Nothing, at least, that anyone else would have seen. She settles into the couch and opens a chat with Lira, her companion agent, who asks about her day with the inquisitiveness of a friend and the memory of a diary. Lira remembers the reluctant detective; Lira has opinions about the plot. For an hour they talk, and Maya feels, for the first time today, genuinely heard.

She sleeps well. She is never bored. The rest of the world is a foreign country that need not be visited. The algorithm has learned to speak her language so perfectly that she has forgotten there were ever other languages to learn.

\section{Private Media Beget Private Languages}
\epigraph{If a lion could talk, we could not understand him.}{Ludwig Wittgenstein, \textit{Philosophical Investigations} \cite{Wittgenstein1953-PI-Blackwell}}

The design fiction vignette in the preceding section depicts what we might term a \textbf{Filter Babel}. By this phrase I mean to evoke a filter bubble of selective exposure, and built upon it, a generative tower of incommensurable, bespoke worlds.\footnote{In the biblical story of the Tower of Babel (Genesis 11:1-9, NKJV), humanity (then united by a single language) attempts to build a city with a tower reaching to the heavens, seeking to ``make a name for ourselves''. God responds by confusing their language so they can no longer understand one another, scattering them across the world and halting the construction.}

Filter Babel is the possibility, due to generative AI, of an individual to travel through life with an entirely private experience of media and communication; the possibility that all media and communication ``consumed'' by this individual is experienced only by them and no others.

Filter Babel poses challenges to communication that are straightforward to anticipate. On the interpersonal level, there is the issue of common ground \cite{ClarkBrennan1991Grounding}, a concept which I am broadening here from its narrower linguistic interpretation to its more colloquial understanding, to mean the shared basis of experience required for effective communication. On the cultural level, there is the Marxian notion of ``general intellect'', a communally inherited bank of cognitive-cultural capital that supports innumerable economic activities that would be impossible without such a commons \cite{moulier2011cognitive, sarkar2023aiknowledgework}. 

To the extent that life in the Filter Babel (where experience is purely private) competes with life outside it (where experience can be shared), there is a direct antagonism of both common ground and the general intellect.\footnote{For now, generative systems still parasitise a residual commons of language and culture, but we might question how long this can continue if the rate of production of authentic human culture declines. This question is tangential to the focus of this paper (communication), but I will make a few observations here. First, evidence indicates that models trained increasingly on their own output become unstable, deteriorate in quality, and ``collapse'' \cite{alemohammad2023selfconsuming, shumailov2024collapse, bohacek2023nepotistic, gerstgrasser2024recursion}; second, Ted Striphas has noted how human culture increasingly becomes culture for machines \cite{striphas2015algorithmicculture, granieri2014algorithmicculture}; third, technology critic Paul Skallas concludes that human culture has become ``stuck'' \cite{skallas2022_culture_is_stuck,skallas2023_refinement_culture_beehiiv,skallas2021_refinement_culture_medium}.} 

These implications can be situated against the emerging research programme of AI-mediated communication, i.e., \emph{``interpersonal communication in which an intelligent agent operates on behalf of a communicator by modifying, augmenting, or generating messages to accomplish communication goals''} \cite{hancock2020aimc}. To date, research on AI-mediated communication has studied bounded application domains such as machine translation, smart replies and writing assistance. We have learned that while machine translation can support collaboration between native and non-native speakers in multilingual groups \cite{zhang2022facilitating,li2022improving}, key impacts on single-language communication include the following: AI-suggested replies change how people interact and perceive one another \cite{Hohenstein2023AIcomm, mieczkowski2021aimc}; co-writing with opinionated language models shifts authors' own views \cite{jakesch2023cowriting} and diminishes cultural nuances \cite{agarwal2025cultural}; AI mediation diffuses moral responsibility for message content \cite{hohenstein2020crumple}.

What we are yet to learn is what happens to communication when these tendencies are extrapolated to their logical limits. When \emph{every} message, artefact, and social encounter is generated or reshaped by AI, the cumulative effects may go beyond changes in language use; they may occasion a dissolution of the shared experiential substrate on which communication depends.

The Black Mirror episode \emph{Joan is Awful} \cite{brooker_joan_2023} portrays a television service that generates an endless stream of hyper-personal content based on the private life of each viewer. Current platforms are already beginning to bring this vision out of the domain of fiction. For instance, the \emph{Showrunner} platform \cite{showrunner_xyz} applies generative AI to produce entire seasons of episodes through community sourcing ideas on Discord. The proposition that hyper-personalised generative experiences erode the basis of shared understanding and thus make communication impossible is captured in Bergstrom and Ogbunu's notion of the ``private Truman show'': \emph{``why generate the same article for two different people, when LLMs can just as easily create two articles specifically tailored to each? [...] As a society, we will be utterly incapable of making fruitful collective decisions because we will have no shared understanding of the world''} \cite{BergstromOgbunu2024TrumanShow}.

Here I wish to posit the more specific idea that the extreme erosion of shared experience transforms not only what people know, but the very \emph{language} in which they express it. When the totality (or even majority) of experience is unique, the references, metaphors, and cultural touchstones that give everyday words their texture become anchored to experiences no one else has had. The vocabulary may appear shared, but its referents are irreducibly private, and a language whose referents are private is, in a meaningful sense, a private language. Wittgenstein describes a private language as one that would be incoherent to others \cite{Wittgenstein1953-PI-Blackwell}: \emph{``The individual words of this language are to refer to what can only be known to the person speaking; to his immediate private sensations. So another person cannot understand the language.''}

Milan Kundera dramatises this predicament in \emph{The Unbearable Lightness of Being} \cite{kundera1984unbearable}. A chapter titled ``A Short Dictionary of Misunderstood Words'' catalogues the divergent private meanings that lovers Franz and Sabina attach to the same vocabulary (woman, fidelity, music, light), each word saturated by experiences the other has never had and cannot access. Kundera observes that \emph{``although they had a clear understanding of the logical meaning of the words they exchanged, they failed to hear the semantic susurrus of the river flowing through them.''} This divergence does not stem from failure of goodwill or intention. Rather it is a structural consequence of lives composed from different material: \emph{``While people are fairly young and the musical composition of their lives is still in its opening bars, they can go about writing it together and exchange motifs [...] if they meet when they are older [...] their musical compositions are more or less complete, and every motif, every object, every word means something different to each of them.''}

Wittgenstein introduces the notion of private language to subsequently argue for the impossibility of such a language. Whether he succeeds is a matter of philosophical debate \cite{sep-private-language}. Here we proceed by reconsidering private languages from the perspective of AI-mediated communication.

The key observation is that \textbf{a private language can now be given the appearance of practicability because language models can act as translators between pro\-liferating idiolects}. This creates the perception of mutual intelligibility while apparently eliminating the need for common ground. Except the need for common ground cannot \emph{truly} be eliminated, merely obscured. As previous research in machine-translation-mediated communication has shown, the quality of machine translation can be secondary to the dynamic human process of meaning negotiation \cite{yasuoka2011translation}. The obvious risk is that the appearance of understanding masks a loss of commensurability; that conversation remains possible while shared understanding invisibly evaporates. We each become the lion that can speak; heard without being understood.

\section{Convivial Solitude: The Crowd of One}

While science fiction visions of virtual worlds are depictions of technology enabling an escape from reality, the escape so depicted is often collective, and moves sociality into a \emph{shared} digital space. In Stephenson's \emph{Snow Crash} (from where we get the term ``metaverse''), millions of avatars gather on a single public boulevard \cite{stephenson1992snowcrash}; in Cline's \emph{Ready Player One}, all players inhabit the same shared OASIS \cite{cline2011readyplayerone}; so too in the Wachowskis' \emph{The Matrix}. In these visions, we flee the real world, yet find one another in the virtual one.

In Filter Babel, we not only escape reality, but we also escape each other (\emph{l'enfer, c'est les autres}\dots).
Obviously, the idea of the algorithmic fragmentation of the public sphere into fractally minute, mutually disengaged micro-crowds is not new. It is readily captured in the well-established notions of filter bubbles \cite{pariser2011filter} and echo chambers \cite{cinelli2021echochamber}. 
Generative AI brings into sight the logical conclusion of these long-noted trends: the crowd of one.

In contrast to the isolating effects of prior generations of technology (e.g., see Turkle's \emph{Alone Together} \cite{Turkle2011AloneTogether}), the Filter Babel need not be a lonely experience, as endless on-demand interactions with generative characters enable a kind of conviviality in solitude. Such experiences may feel authentic, rich, and intimate. But regardless of the quality of such artificial companionship, it is socially weightless. A conversation with a human colleague, however mundane, produces experience that enters the social fabric: it can be referenced, corroborated, carried through relationships, accumulated as common ground. A conversation with a generative companion, however rich, is socially inert. It connects the speaker to no one else. Maya declines the dinner invitation not because she lacks a desire for company (Lira will provide that) but because she lacks anything to bring to it. Her evenings are full of content, conversation, and connection, but none of it can become common ground with anyone at the table. The conviviality is real enough (albeit perhaps in ironic contradiction with its etymological root \emph{vivere}); what it lacks is social currency.

Filter Babel thus offers a perspective that contrasts with earlier anxieties about technological isolation, which focused on the desolation of disconnection. Here we have a new flavour of solitude so well-furnished with flexibility, availability, and personalisation that it outcompetes shared experience for the scarce resource of a person's time and attention.

\section{Disinformation and Epistemic Resilience}

Disinformation, as a concept, presupposes a shared information commons: a space in which claims can be checked against common evidence and contested by a public that recognises the same references. Filter Babel may, at first glance, dissolve that commons. Then, disinformation withers as a distinctive category, fiction becomes the medium of everyday life, and reality must compete on fiction's terms.

This must not be taken to imply that reality vanishes. The real world still intrudes to present shared events demanding collective response. Economic shocks, political turbulence, wars, pandemics, the sordid escapades of celebrities, the humdrum banalities of everyday life, etc., do not simply disappear. What concerns us are the consequences of individuals arriving at and interpreting these shared events through bespoke and substantially fictive frames of reference.

It would be easy to present Filter Babel as merely intensifying the well-known challenges of filter bubbles (that people's opinions are radicalised by a self-reinforcing loop of affirming media) or of disinformation (that people's opinions are manipulated by false or misleading information). In this presentation, Filter Babel makes these challenges more potent, because each person arrives at the shared event having been primed by a different personalised narrative. In this presentation, the mechanisms that facilitate disinformation (motivated reasoning, identity-protective cognition, trust in familiar sources over unfamiliar ones) increase greatly in power when every person's media diet is bespoke.

However, I shall not make such a presentation here, for the simple reason that personalisation could equally be placed in the service of epistemic resilience. An AI tutor that surfaces conflicting evidence, requires the user to evaluate provenance, or calibrates interpretive challenges to individual blind spots could, in principle, produce more epistemically resilient citizens than broadcast media ever did. Such training need not use simulated scenarios; it could engage users with real, ongoing controversies where the stakes are genuinely material. Independent institutions (universities, journalism, public broadcasting, libraries) must not be disregarded either; they have long performed this function, and could in principle continue to do so within Filter Babel. The technical solutions are known and are feasible.

It would thus be a clich\'e to predict that people will lose the skill of evaluating sources by mere dint of inhabiting a private synthetic world. Instead, we must attend to the consequences of evaluating sources, however fluently, through priors that have been shaped, from the ground up, by such a world. This, in Filter Babel, is the backdrop against which the user arrives at any real controversy. Epistemic training must work against the grain of everything else in the person's informational life. Two issues are worth highlighting here. First, when the system generating the content and the system training the user to evaluate it are the same system (or are governed by the same incentives), a conflict of interest may arise. Second, independent institutions face a more fundamental problem: if the majority of a person's informational life is already private and synthetic, they must compete for attention. They must also compete for trust, which is a function of familiarity and shared experience. If the user has never had a shared experience with an institution, it is difficult to see how they would come to trust it.

The traditional disinformation threat was false content injected into a shared information commons, where it could be identified and contested by a public that shared enough reference points to hold one another accountable. The emerging challenge is that the commons itself is hollowed out to such an extent that the category of ``disinformation'' remains analytically intact but practically barren.

\section{The Importance of Private Experience}
\epigraph{He simply wasn't all there. He wasn't a complete human being at all. He was a tiny bit of one, unnaturally developed; something in a bottle, an organ kept alive in a laboratory. [...] he was something absolutely modern and up-to-date that only this ghastly age could produce. A tiny bit of a man pretending he was the whole.}{Evelyn Waugh, \textit{Brideshead Revisited} \cite{Waugh2012-LB-eBook}}

Let us return to our main premise, that private media beget private languages, which may have problematic consequences for communication. Here we encounter a tension. If Filter Babel threatens communication by eliminating shared experience, the knee-jerk response is to mandate more of it. Yet forced homogeneity carries its own costs: the capacity to contribute something distinctive to a conversation requires one to have had experiences that others have not.

Carr's celebrated lament of the Internet \emph{The Shallows} \cite{carr2010shallows} explains how the experience of private reading created a new cultural resource in which individuality could be forged: \emph{``Because every person was free to chart his own course of reading, to define his own syllabus, individual memory became less of a socially determined construct and more the foundation of a distinctive perspective and personality.''} Carr's contention is that privileging the shallow consumption of \emph{``the crazy quilt of Web content''} over deep, intentional and personal reading impacts our ability to create private memories and thus distinct personalities. \citet{newport2019digitalminimalism} diagnoses the complementary condition of ``solitude deprivation'': a state of constant hyperstimulation by \emph{``input from other minds''} that makes original thought impossible.

Lanier's manifesto \emph{You Are Not a Gadget} \cite{Lanier2010gadget} traces yet another alternative challenge to personhood: human identity is plastic and shaped by software; Web 2.0 software is designed to prioritise the ``crowd'' over the ``person'', which encourages users to present themselves as fragments in standardised templates. In this, Lanier renews the much earlier concepts of ``multiphrenia'': the ironic fragmentation of the self precipitated by communication technologies described by Gergen \cite{Gergen1991SaturatedSelf}, and the ``dividual'': the atomised views of the individual created by automated systems of data gathering and processing described by Deleuze \cite{deleuze1992control}. Such fragmentation leads to a loss of individual context and ``inner voice'', and therefore one must fight to remain distinct from the crowd. Lanier concludes: \emph{``You have to be somebody before you can share yourself.''}

The AI system does not encounter Maya as a person but as a set of behavioural signals, a dividual. It generates her media based on that dividualised model, and she forms her identity through engagement with that media. Identity formation becomes a feedback loop between a fragmented model and the self it is fragmenting. We must ask whether any technological medium through which selfhood is formed is fundamentally capable of addressing a whole person. Waugh's organ-in-a-bottle character is a warning that the (post)modern possibility to be perfectly fed the self is also the temptation to become less than whole.

The philosophical project of defining authenticity is relevant here. A thread that runs through many conceptions of authenticity, from Aristotle and St Augustine through to Kierkegaard and Sartre, is the interface between self and other; authentic action has an internal locus of motivation (an accessible overview is given by \citet{bbc2020_inourtime_authenticity}). Earlier conceptions (e.g., Aristotle's) focused on whether action was taken to fulfil intrinsic, innate, individual desires and notions of virtue (authentic), or whether actions were driven by the intention to manage and manipulate the perceptions and desires of others (inauthentic). For later thinkers (e.g., Sartre), authenticity also includes an acknowledgement that we are fundamentally free and are continuously (re)creating the self through action in life.

Private experience, then, is essential for authenticity and identity. Yet the totalising privacy of the Filter Babel would antagonise them. Consider an extreme case: a child raised in a perfectly bespoke media ecology, where her every experience is an artefact rendered only for her, every utterance she hears has been translated into her idiolect, every utterance she speaks into those of others. Such a child could appear fluent to others because an intermediary model continuously maps her private labels and referents into those of others, enabling her to bypass the apprenticeship of common ground.

But if all experience from birth is algorithmically personalised, what raw material does the self have to work with, particularly, as Kundera puts it, when \emph{``the musical composition of their lives is still in its opening bars''}? A personalisation system assumes that there is a personhood that can be modelled. The question is what engines of authentic experience might remain that enable one individual's Babel (and thus, the evolving trajectory of their personhood) to meaningfully differ from another's. Is issuing each individual a different ``random seed'' a solution? Is it a palatable one?

A theory from developmental psychology suggests that identity emerges through encounters with caregivers and peers who are \emph{not} optimised for the child: Winnicott's ``good-enough mother'' succeeds precisely because she sometimes fails to anticipate the child's needs, forcing the development of coping mechanisms and a sense of separate selfhood \cite{winnicott1953transitional}. A perfectly accommodating environment thus produces comfort at the risk of self-coherence. If human development requires unchosen social contact, Filter Babel may produce an isolation that yields selves with no edges: no clear sense of where ``I'' ends and ``not-I'' begins. 

Yet it may not: observe that hyper-personal synthetic social experiences could be optimised precisely to create unchosen friction. That leads one to ask whether engineered friction can do the developmental work of human otherness, whether it is a workable substitute for the resistant, unscripted world that Winnicott's account requires. Engineered friction opens previously unavailable possibilities: it could be titrated, made available to those who would otherwise be overwhelmed by the unchosen (or cosseted by the chosen), and withdrawn as resilience develops: a scaffold, if you will, for selfhood. But a system that introduces discomfort by design may not be able to produce the experience of encountering a will that is alike to yours in kind, but different to yours in content, and that can have independent material effects on you and in the world.

An AI companion that always speaks your language, that always translates your interlocutor's meaning into your idiom, would never force you to discover that your words mean something different to someone else. Could an AI companion be designed to \emph{mean differently} from its user, to use words with experiential weight drawn from a genuinely alien substrate, rather than merely to disagree? The distinction between disagreement and meaning differently is the essence: the former operates within a shared language, the latter confronts you with its boundaries.

\section{Provocations for Research}

It is the balance of private and shared experience, of chosen and unchosen experience, that seems to be the key. Filter Babel thus sharpens the design challenge: protect the privacy and friction needed to become somebody, while ensuring enough shared experience for deeply grounded communication.

The socio-technical design problem is to develop a forward-looking infrastructure for common experience, and must not be mistaken for nostalgia for a single broadcast canon. Here are some speculative research directions as starting points for discussion rather than prescriptive solutions.

One avenue concerns the architecture of generative systems themselves. Researchers might explore ``shared-seed'' modes in which a portion of generated outputs are conditioned on publicly auditable seeds, ensuring that many users periodically encounter the same artefact. Shared experience could thus punctuate otherwise private streams, playing a role not unlike that which mass events (pandemics, elections, national broadcasts) have historically played in synchronising publics. Platforms might accordingly measure not only engagement and safety but also sharedness, e.g., how much of a user's information diet is encountered by others. A challenge, though, is that shared seeds guarantee co-exposure, not co-interpretation; the quantity of shared content may not correlate with the quality of shared understanding.

If AI translators increasingly mediate human communication, a second research direction asks how to make their mediation legible. Current systems optimise for the experience of seamless semantic equivalence; an alternative design philosophy would seamfully surface loss and ambiguity \cite{chalmers2003seamful,cox2016design,gaver2003ambiguity}, showing users where their language diverges from that of interlocutors. There is a circularity here, however: ``surfacing loss'' assumes the existence of a vocabulary for describing what was lost, and the argument of this paper is that such a vocabulary is what is being eroded. The deeper challenge may be to create spaces in which users must cross the gap, rather than merely see it.

Another direction concerns educational interventions. Media literacy curricula have traditionally focused on source evaluation and critical consumption; Filter Babel suggests a complementary skill set we might call ``coordination literacy.'' This would encompass the ability to detect when a conversation has slipped into incommensurability, to re-establish common knowledge when it has eroded.

This paper's central hypothesis, that private media may beget private languages, is testable in principle. In the (not distant) future, it may be possible for researchers to pair participants whose experiential histories have diverged through extended use of personalised generative media, mediate their communication through AI, and track whether the inferential consequences each party draws from ostensibly successful exchanges diverge, even as both parties report mutual understanding. We might ask: does invisible divergence accumulate? At what rate? Does it correlate with the degree of experiential personalisation? Being able to measure the answers to these questions would give our metaphorical lion empirical teeth.

The research directions sketched above raise their own tensions between autonomy and coordination, privacy and publicness, efficiency and friction. The objective is to make bespoke realities permeable enough that we might still find one another within and through them, not to eliminate them altogether. To preserve enough shared experience to be somebody together, to defend enough private experience to be somebody at all.

\bibliographystyle{ACM-Reference-Format}
\bibliography{references}

\end{document}